\documentclass[fleqn,10pt]{wlscirep}
\usepackage[utf8]{inputenc}
\usepackage[T1]{fontenc}
\usepackage{upgreek}
\usepackage{soul,xcolor}
\usepackage{ulem}

\title{Multiband effects on the upper critical field angular dependence of 122-family iron pnictide superconductors}

\author[1]{I. F. Llovo}
\author[1]{C. Carballeira}
\author[1]{D. S\'o\~nora}
\author[1]{A. Pereiro}
\author[2]{J. J. Ponte}
\author[3]{S. Salem-Sugui Jr.}
\author[4]{A. S. Sefat}
\author[1,*]{J. Mosqueira}

\affil[1]{QMatterPhotonics Research Group, Departamento de F\'isica de Part\'iculas, Universidade de Santiago de Compostela, E-15782 Santiago de Compostela, Spain}

\affil[2]{Unidade de Magnetosusceptibilidade, RIAIDT, Universidade de Santiago de Compostela, E-15782 Santiago de Compostela, Spain}

\affil[3]{Instituto de Fisica, Universidade Federal do Rio de Janeiro, 21941-972 Rio de Janeiro, RJ, Brazil}

\affil[4]{Oak Ridge National Laboratory, Oak Ridge, TN87831, United States of America}

\affil[*]{j.mosqueira@usc.es}

\begin{abstract}
\setstcolor{red}
Detailed measurements of the in-plane resistivity were performed in a high-quality Ba(Fe$_{1-x}$Co$_{x}$)$_2$As$_2$ ($x=0.065$) single crystal, in magnetic fields up to 9 T and with different orientations $\theta$ relative to the crystal $c$ axis. A significant $\rho(T)_{H,\theta}$ rounding is observed just above the superconducting critical temperature $T_c$ due to Cooper pairs created by superconducting fluctuations. These data are analyzed in terms of a generalization of the Aslamazov-Larkin approach, that extends its applicability to high reduced-temperatures and magnetic fields. This method allows us to carry out a criterion-independent determination of the angular dependence of the upper critical field, $H_{c2}(\theta)$. In spite of the relatively small anisotropy of this compound, it is found that $H_{c2}(\theta)$ presents a significant deviation from the single-band 3D anisotropic Ginzburg-Landau (3D-aGL) approach, particularly for large $\theta$ (typically above $\sim60^o$). These results are interpreted in terms of the multiband nature of these materials, in contrast with other proposals for similar $H_{c2}(\theta)$ anomalies. Our results are also consistent with an \textit{effective} anisotropy factor almost temperature independent near $T_c$, a result that differs from the ones obtained by using a single-band model.
\end{abstract}
\begin{document}

\flushbottom
\maketitle
%
%
\thispagestyle{empty}

\section*{Introduction}
\setstcolor{red}

Since the discovery of superconductivity at relatively high temperatures in Fe-based superconductors (FeSC)\cite{Kamihara08} in 2008, intensive research on these materials has been taking place. On the one hand, these materials present high critical magnetic fields and low anisotropies, for which they have received great attention towards their potential applications in electric transport under high magnetic fields.\cite{Hosono18,Pyon20} On the other hand, there is a fundamental interest in discovering the pairing mechanism responsible for their high critical temperature, which could be related to the one of cuprates. They also present unconventional superconducting properties associated to their multiband electronic structure, with energy gaps that depend on the doping level and on the pressure (external or chemical).\cite{Paglione10,Hirschfeld11,Wang11,Stewart11,Chen14} An example of this would be the anomalous temperature dependences of the magnetic penetration depth,\cite{Prozorov11} the specific heat,\cite{Hardy10,Maksimov11} or the upper critical field,\cite{Hunte08,Gurevich11,Xing17} that have been interpreted in terms of theoretical models with two effective superconducting gaps.\cite{Gurevich03,Gurevich10}

The angular dependence of the upper critical magnetic field, $H_{c2}(\theta)$, where $\theta$ is the angle between the applied magnetic field and the crystal $c$ axis, has been less studied. If the bands contributing to the superconductivity have different anisotropies, then $H_{c2}(\theta)$ may differ from the single-band 3D-anisotropic Ginzburg-Landau (3D-aGL) approach, that may be written as\cite{Tinkham}
\begin{equation}
H_{c2}(\theta)=\left(\frac{\cos^2\theta}{H_{c2}^{\perp2}}+\frac{\sin^2\theta}{H_{c2}^{\parallel2}}\right)^{-1/2},
\label{refeq1}
\end{equation}
where $H_{c2}^\perp$ and $H_{c2}^\parallel$ correspond to $\theta=0^\circ$ and $90^\circ$ respectively.
In some compounds from the 122 family $H_{c2}(\theta)$ is well described by equation~(\ref{refeq1}), with a moderate anisotropy factor $\gamma\equiv H_{c2}^\parallel/H_{c2}^\perp$ \cite{Terashima09,Yuan09,Su14,Hao15} However, an anomalous behavior has been reported in some of these compounds,\cite{Murphy13,Hanisch15} qualitatively similar to the behavior observed in other two-band superconductors, such as MgB$_2$.\cite{Shi03,Kim06} 

The reason for the discrepancies lies in the experimental difficulties to determine the upper critical field. The $H_{c2}(T)$ line is generally obtained from the $T$ and $H$ pairs at which the electrical resistivity falls to a given percentage of the normal-state resistivity.\cite{Terashima09,Yuan09,Su14,Hao15,Murphy13,Hanisch15} Nonetheless, this procedure is strongly dependent on the particular criterion used (e.g., 20\%, 50\% or 80\% of $\rho_n$), as different factors \textit{round} the $\rho(T)$ curves near the superconducting transition temperature, $T_c(H)$. Firstly, due to the relatively high $T_c$ and the small value of the coherence length (just a few nm), thermal fluctuation effects near $T_c(H)$ play an important role in these materials,\cite{Mosqueira11} and contribute to the rounding of the resistive transition.\cite{Rey13,Ramos15,Sonora17,Ahmad17,Ahmad18} These effects are also strongly dependent on the amplitude and orientation of the applied magnetic field.\cite{Rey14} 
A second factor is the effect of $T_c$ inhomogeneities: These compounds are generally non-stoichiometric, and their $T_c$ depends on the doping level. Given the small values of the coherence length, it is expected that even a random distribution of dopants would lead to nanoscale $T_c$ variations,\cite{Cabo09} which results in a smoothing of the resistive transition, an effect particularly important in non-optimally-doped compounds.\cite{Rey13} Finally, the resistive transition is extended by vortex dynamics below $T_c(H)$, down to the irreversibility temperature, under which the vortices are pinned.

In this work, we present measurements of the in-plane resistivity versus temperature under magnetic fields with different amplitudes and orientations with respect to the crystal $c$ axis, in an optimally-doped Ba(Fe$_{1-x}$Co$_x$)$_2$As$_2$ (OP-BaFeCoAs) single crystal. This compound is one of the most studied FeSC, and nowadays it is possible to grow large single crystals of both the highest stoichiometric and structural quality. 
In contrast with the aforementioned procedures to obtain the upper critical field, the $\rho(T)_{H,\theta}$ rounding will be studied in terms of a generalization of the Aslamazov-Larkin (AL) approach for the effect of superconducting fluctuations, which is applicable in the region of high reduced temperatures and magnetic fields.\cite{Rey13} The analysis will allow us to obtain a criterion-independent determination of the angular dependence of the upper critical field, $H_{c2}(\theta)$. The result will be analyzed in terms of existing models for multiband superconductors.\cite{Gurevich03,Gurevich10}

\section*{Results}

\begin{figure}[ht]
\centering
\includegraphics[width=\linewidth]{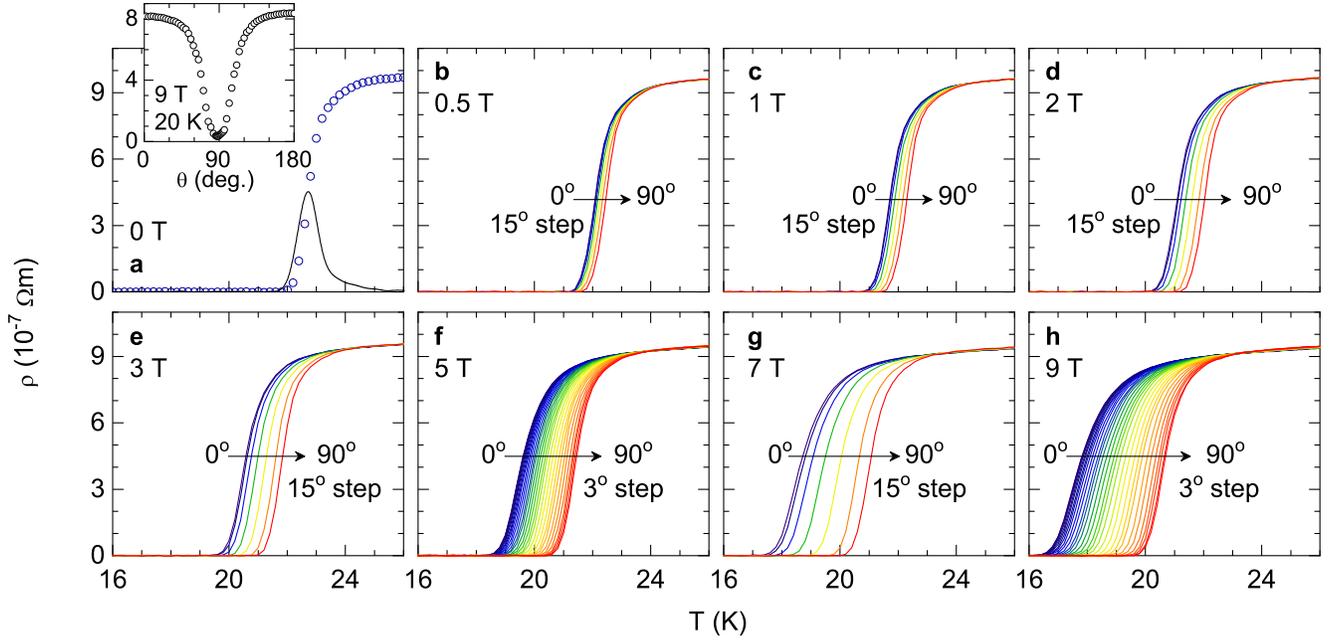}
\caption{Temperature dependence of the in-plane resistivity around $T_c$ in the presence of magnetic fields with different amplitudes (from 0 to 9 T, \textbf{a} to \textbf{h}), and different orientations with respect to the crystal $c$ axis. $T_c=22.7$~K was determined as the temperature at which $(d\rho/dT)_{H=0}$ is maximum (solid line in \textbf{a}). Inset in \textbf{a}: $\rho(\theta)$ measurement performed before the measurements in \textbf{a-h} to determine the precise $\theta=90^\circ$ position.}
\label{reffig1}
\end{figure}

Figure~\ref{reffig1} shows the resulting $\rho(T)_{H,\theta}$ near the superconducting transition temperature in detail. The zero-field transition temperature $T_c=22.7$~K was estimated from the maximum of the $d\rho/dT$ curve (solid line in Fig.~\ref{reffig1}\textbf{a}). The $T_c$ uncertainty is $\pm 0.5$~K, which is primarily caused by the resistivity rounding associated with superconducting fluctuations.\cite{Rey13} As it can be seen, $T_c$ shifts to lower temperatures as the magnetic field is increased. This effect is more pronounced for $H\perp ab$ (i.e., $\theta=0$) than for $H\parallel ab$ ($\theta=90^\circ$), due to the anisotropy of the corresponding upper critical fields, $H_{c2}^\perp$ and $H_{c2}^\parallel$, respectively. 
These data may be then used to estimate $H_{c2}(T,\theta)$. However, in the available range of magnetic fields, the $T_c$ shift is close to the $T_c$ uncertainty, mainly attributed to the aforementioned resistivity rounding. For this reason, the results are highly dependent on the criterion used to determine $T_c(H,\theta)$ (typically the temperature at which the resistivity falls to a given fraction of the extrapolated normal-state resistivity). In the next section, a criterion-independent determination of the angular dependence of $H_{c2}$ will be presented through the analysis of the superconducting fluctuations, obtained from the rounding above $T_c(H,\theta)$.

\subsection*{Determination of the normal state background}

The conductivity induced by superconducting fluctuations (or paraconductivity) is given by
\begin{equation}
\Delta\sigma(T)_{H,\theta}=\frac{1}{\rho (T)_{H,\theta}}-\frac{1}{\rho_B (T)_{H,\theta}}
\label{refeq2}
\end{equation}
where $\rho_B (T)_{H,\theta}$ is the normal-state or background resistivity extrapolated to temperatures near $T_c$. This background resistivity was determined by a linear fit to the resistivity above 35~K (i.e., above $1.5T_c$) where fluctuation effects are expected to be negligible.\cite{Rey13,Ramos15,Sonora17,Ahmad17,Ahmad18} Some examples of this procedure for different field amplitudes and orientations are presented in Figs.~\ref{reffig2}\textbf{a-c}. 

\subsection*{Analysis of $\Delta\sigma$ for $\theta=0$ data}

\begin{figure}[ht]
\centering
\includegraphics[width=0.9\linewidth]{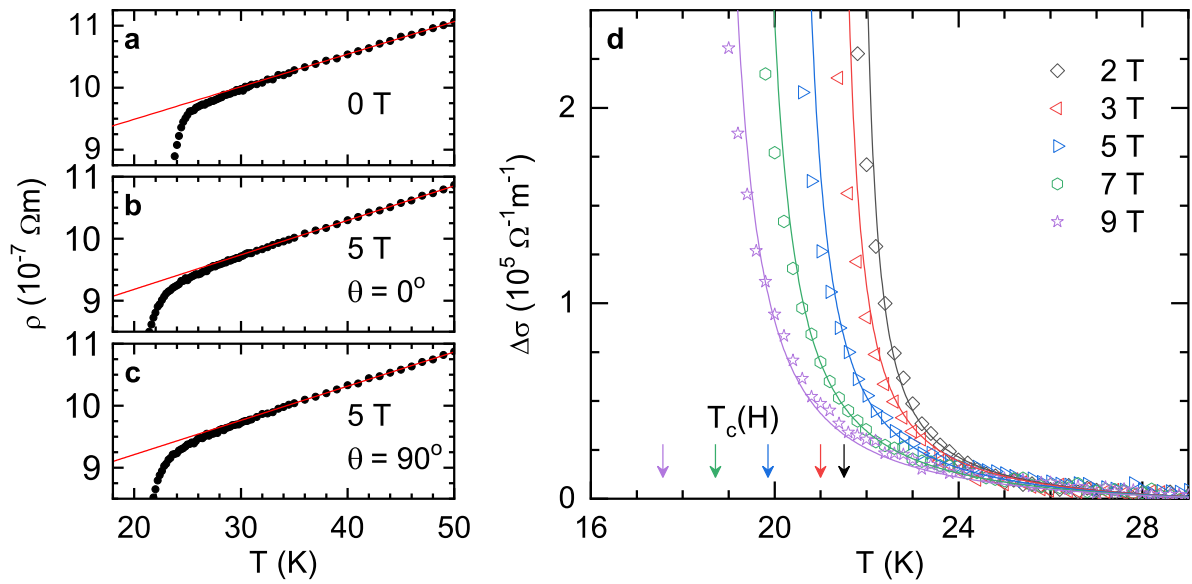}
\caption{\textbf{a-c} Some examples (for different $H$ amplitudes and orientations) of the in-plane resistivity temperature dependence well above $T_c$. The normal-state backgrounds (red lines) were determined by linear fits above 35~K ($\sim1.5T_c$), where fluctuation effects are negligible. \textbf{d} Temperature dependence of the fluctuation conductivity under different magnetic fields, applied perpendicular to the crystal $ab$ layers ($\theta=0$). The lines are the best fit of equation~(\ref{refeq3}) to the data between 2 and 9 T. The arrows indicate $T_c(H)=T_c(1-H/H_{c2}^\perp(0))$ for the magnetic fields used in the experiments.  }
\label{reffig2}
\end{figure}

A first comparison with the experimental data was performed for the $\Delta\sigma$ data for $\theta=0$ presented in Fig.~\ref{reffig2}\textbf{d}. As it can be seen, the rounding associated with fluctuation effects can be clearly observed a few degrees above $T_c$. The data were analyzed in terms of the 3D-anisotropic Ginzburg-Landau (GL) approach developed in Ref.~\citeonline{Rey13}, which is valid under finite magnetic field amplitudes. For $H\perp ab$, it may be written as
\begin{equation}
\Delta\sigma(\varepsilon,h)=\frac{e^2}{32\hbar \pi\xi_c(0)}\sqrt{\frac{2}{h}}\int_0^{\sqrt{\frac{c-\varepsilon}{2h}}}{\rm d}x
\left[\psi^1\left(\frac{\varepsilon+h}{2h}+x^2\right)-\psi^1\left(\frac{c+h}{2h}+x^2\right)
\right],
\label{refeq3}
\end{equation}
where $\varepsilon\equiv\ln(T/T_c)$ is the reduced temperature, $h\equiv H/H_{c2}^\perp$ the reduced magnetic field, $H_{c2}^\perp$ the linear extrapolation to $T=0$~K of the upper critical field for $H\perp ab$, $e$ the electron charge, $\xi_c(0)$ the $c$ axis coherence length amplitude, and $c$ a cutoff constant of the order of magnitude of the unity, introduced to exclude the contribution of high-energy fluctuation modes.\cite{Vidal02} It is clear to see that $c$ corresponds to the reduced temperature at which fluctuation effects vanish. As it can be seen in Figs.~\ref{reffig2}\textbf{a-c}, the measured $\rho(T)$ deviates from $\rho_B(T)$ (beyond the experimental uncertainty) when $T<30-31$~K, which corresponds to a reduced temperature around 0.3. Thus, in what follows we have set $c=0.3$, a value that is close to the one found in other FeSC.\cite{Rey13,Ramos15,Sonora17,Ahmad17,Ahmad18} In the zero-field limit (for $h\ll\varepsilon$), and in the absence of cutoff ($c\to \infty$), equation~(\ref{refeq3}) reduces to the well known Aslamazov-Larkin expression, $\Delta\sigma(\varepsilon)=e^2/32\hbar\xi_c(0)\varepsilon^{1/2}$. 

The lines in Fig.~\ref{reffig2}\textbf{d} are the best fit of equation~(\ref{refeq3}) to the set of data obtained with fields between 2 and 9~T, with only two free parameters: $\xi_c(0)$, which is directly related to the $\Delta\sigma$ amplitude, and $H_{c2}^\perp$, which is implicit in the equation through the reduced magnetic field $h$, and which is related to the temperature shift of $\Delta\sigma$ induced by the magnetic field. As it can be seen, the agreement is excellent, leading to $\xi_c(0)=6.89\pm0.15$~\r{A} and $H_{c2}^\perp=42.5\pm0.5$~T. Experimental data up to 1~T were excluded from the fitting as a significant disagreement with the GL approach was found in previous works, while studying the fluctuation effects in other FeSC families. It has been hypothesized that this discrepancy may arise from a $T_c$ distribution\cite{Rey13,Ramos15} or from phase fluctuations,\cite{Bernardi10, Prando11,Bossoni14} which could be relevant near $T_c$ and under low fields in these materials.

\subsection*{Analysis of $\Delta\sigma$ for arbitrary $\theta$ and angular dependence of $H_{c2}$}

\begin{figure}[th]
\centering
\includegraphics[width=.8\textwidth]{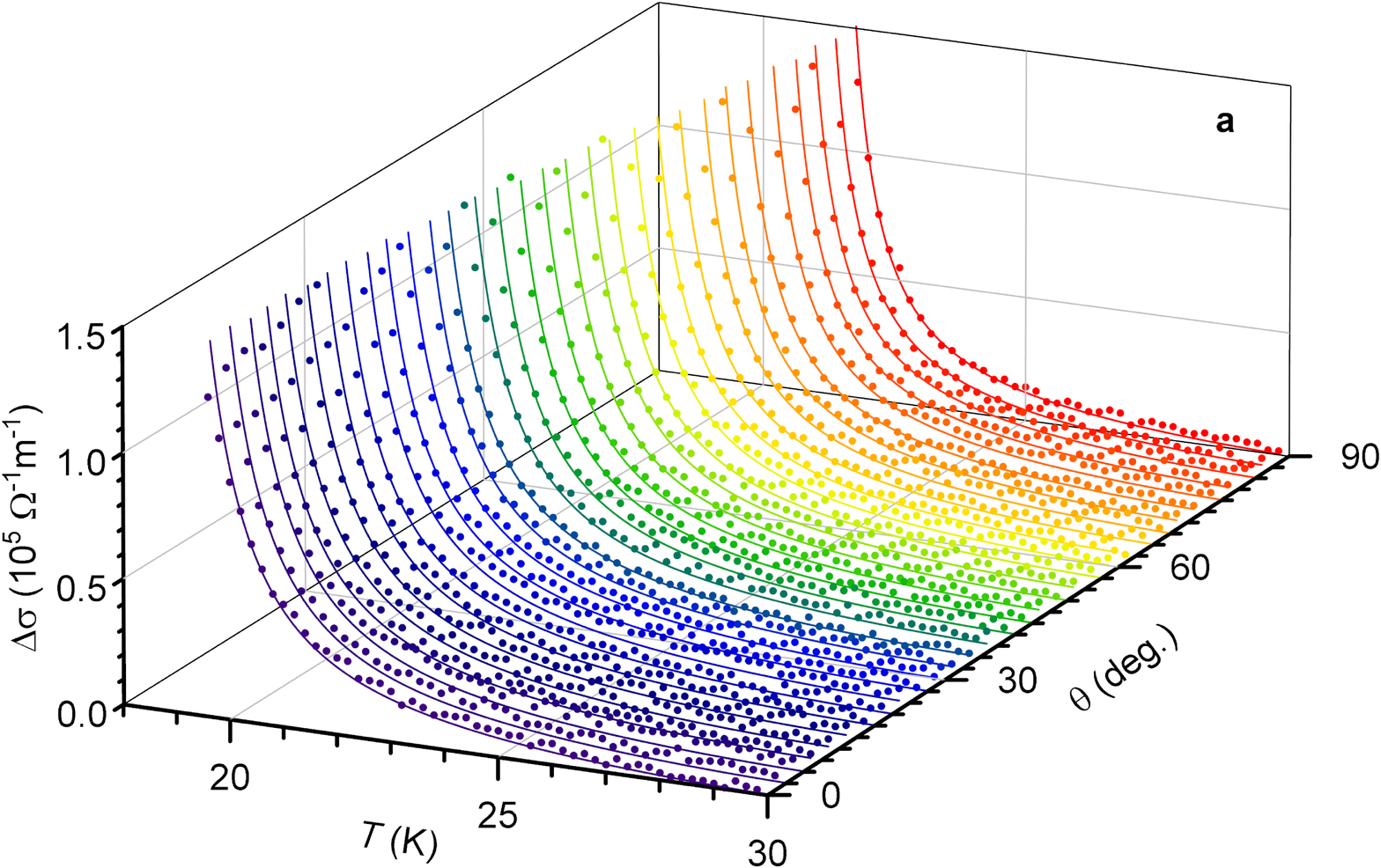}\\ \vspace{5pt}
\includegraphics[width=\textwidth]{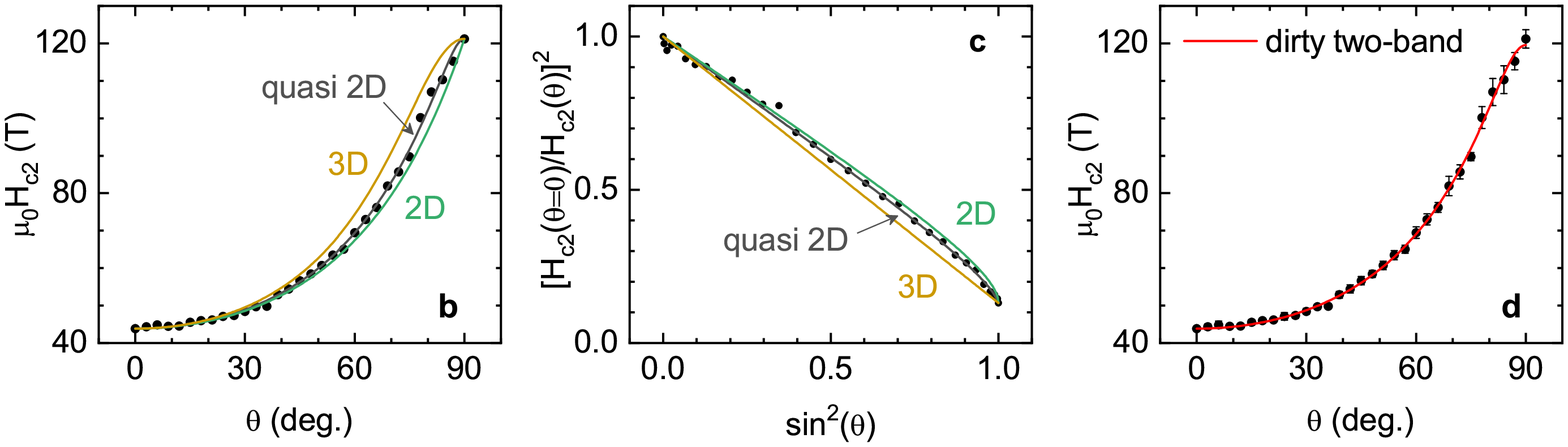}
\caption{\textbf{a} Temperature dependence of the fluctuation conductivity obtained with a 9~T magnetic field, for different orientations relative to the crystal $c$ axis ($\theta$-steps of 3$^\circ$ between 0 and 90$^\circ$). The lines are the best fits of equation~(\ref{refeq3}), with the $\xi_c(0)$ value resulting from the analysis in Fig.~\ref{reffig2}\textbf{d}, and $H_{c2}(\theta)$ as the only free parameter. The resulting $H_{c2}(\theta)$ are the data points in \textbf{b}, \textbf{c} and \textbf{d} (the error bars are only shown in \textbf{d} for clarity). \textbf{b} $H_{c2}(\theta)$ data compared to the single-band 3D-anisotropic GL approach, (orange line, equation~(\ref{refeq1})), the 2D Tinkham's result (green line, equation~(\ref{refeq5})) and quasi-2D Mineev's result (black line, equation~(\ref{refeq6})). \textbf{c} Same plot as \textbf{b}, in a linearized scale. \textbf{d} $H_{c2}(\theta)$ compared with Gurevich's approach for dirty 2-band superconductors (equation~(\ref{GurevichHc2})).}
\label{reffig3}
\end{figure}

We will now analyze the experimental data obtained with different $H$ orientations. To this purpose, the reduced magnetic field in equation~(\ref{refeq3}) must be replaced by\cite{Rey14}
\begin{equation}
h=\frac{H}{H_{c2}(\theta)},
\label{refeq4}
\end{equation}
where $H_{c2}(\theta)$ is the upper critical field (linearly extrapolated to $T=0$~K) for an arbitrary field orientation relative to the $c$ axis. 
The $\Delta\sigma(T)$ data in Fig.~\ref{reffig3}\textbf{a} were obtained under a 9~T magnetic field applied with different orientations ($\theta$ runs from 0 to 90$^\circ$ in steps of 3$^\circ$). The lines are fits of equation~(\ref{refeq3}) to each $\theta$ dataset with the above $\xi_c(0)$ and $c$ as fixed parameters, and $H_{c2}(\theta)$ as the only free parameter. As it can be seen, the fits are in excellent agreement with our data. The resulting angular dependence of the upper critical field is presented in Figs.~\ref{reffig3}\textbf{b, c, d}. 
From this figure, it follows that the upper critical fields extrapolated to $T = 0$~K are 43~T for $H\perp ab$ and 120~T for $H\parallel ab$. The corresponding slopes at $T_c$, -1.9~T/K for $H\perp ab$  and -5.3~T/K for $H\parallel ab$, are close to the ones found in the literature.\cite{Ni08,Sun09,Tanatar09,Yamamoto09,Kim10,Vinod11}
The orange line in this figure is the prediction of the single-band 3D anisotropic GL approach, equation~(\ref{refeq1}), evaluated with the experimental $H_{c2}^\perp$ and $H_{c2}^\parallel$. A good agreement is found at low $\theta$, but for large $\theta$ the behavior is qualitatively closer to the one found in 2D superconductors, approaching 90$^\circ$ with a finite slope. This can be clearly seen in the linearized representation shown in Fig.~\ref{reffig3}\textbf{c}.
For comparison, Tinkham's result\cite{Tinkham63,Tinkham68} for the upper critical field of 2D superconductors evaluated with the experimental $H_{c2}^\perp$ and $H_{c2}^\parallel$ has been included in Figs.~\ref{reffig3}\textbf{b} and \ref{reffig3}\textbf{c} (green line)
\begin{equation}
\left|\frac{H_{c2}(\theta)\cos\theta}{H_{c2}^\perp}\right|+ \left(\frac{H_{c2}(\theta)\sin\theta}{H_{c2}^\parallel}\right)^2=1.
\label{refeq5}
\end{equation}
As it can be seen, the experimental data fall between the 3D and 2D approaches. $H_{c2}(\theta)$ for layered quasi-2D superconductors was obtained in Ref.~\citeonline{Mineev01}, and reads 
\begin{equation}
\frac{H_{c2}^2(\theta)\sin^2\theta}{H_{c2}^{\parallel 2}}\left(1-\frac{H_{c2}^\parallel}{\gamma H_{c2}^\perp}\right)+\frac{H_{c2}(\theta)}{H_{c2}^\perp}\sqrt{\cos^2\theta+\frac{\sin^2\theta}{\gamma^2}}=1,
\label{refeq6}
\end{equation}
where $\gamma=(m^*_c/m^*_{ab})^{1/2}$ is the anisotropy factor. This expression reduces to equations~(1) and (5) in the appropriate limits, and fits the data in Fig.~\ref{reffig3}\textbf{b} in the entire $\theta$-range with $\gamma$ as free parameter (and by setting $H_{c2}^\perp$ and $H_{c2}^\parallel$ to the experimental values). However the resulting $\gamma$ (that in this model is different from the ratio $H_{c2}^\parallel/H_{c2}^\perp$) is as high as 16.5, which is abnormally large for this compound, and inconsistent with the 3D nature of $\Delta\sigma$ in the whole temperature range above $T_c$. In comparison, a value of $\gamma\sim10$  is found in optimally-doped YBa$_2$Cu$_3$O$_{7-\delta}$, and fluctuation effects already present a 3D-2D crossover at temperatures relatively close to $T_c$ ($\varepsilon\sim10^{-1}$).\cite{Tinkham} This indicates that the excellent fit of equation~(\ref{refeq6}) is spurious, and that the anomalous angular dependence of $H_{c2}(\theta)$ cannot be attributed to a quasi-2D behavior. 

Another possibility is that the anomalous $H_{c2}(\theta)$ behavior arises from the multiband nature of these materials. The presence of two effective superconducting gaps in Ba(Fe$_{1-x}$Co$_x$)$_2$As$_2$ was revealed by angle-resolved photoemission spectroscopy (ARPES),\cite{Terashima09b} and point-contact Andreev reflection.\cite{Tortello10} Theoretical models for two-band superconductors also accounted for the anomalous temperature dependence of the magnetic penetration depth, \cite{Williams09, Choi10, Fischer10, Gordon10, Luan10, Luan11, Maksimov11, Yong11} and of the specific heat\cite{Maksimov11, Hardy10} in OP-BaFeCoAs. A recent review on the relevance of multiband effects in Fe-based and other superconductors may also be seen in Ref.~\citeonline{Milosevic15}. However, it is worth noting that, in some cases, multiple superconducting bands and anisotropy affect some observables similarly (see e.g., Ref.~\citeonline{Bekaert16} on the anomalous $T$-dependence of the superfluid density of OsBe$_2$). Nonetheless, our previous analysis clearly shows that the $H_{c2}(\theta)$ dependency cannot be explained with a reasonable anisotropy factor.

The angular dependence of the upper critical field in two-band superconductors was calculated by Gurevich in both the dirty\cite{Gurevich03} and clean limits.\cite{Gurevich10,Gurevich11} A criterion for a superconductor to be in the dirty limit may be expressed as $\hbar/\pi\Delta(0)\gg\tau$, where $\Delta(0)$ is one-half the superconducting energy gap at $T=0$~K, and $\tau$ the quasiparticles relaxation time. In OP-BaFeCoAs the small and large gaps are, respectively, $\sim3k_BT_c$ and $\sim6k_BT_c$ (see e.g., Refs.~\citeonline{Luan10,Luan11,Williams09,Fischer10,Hardy10,Tortello10,Terashima09b}), which leads to \linebreak $\hbar/\pi\Delta(0)\sim(7-3.5)\times10^{-14}$~s. In turn, near $T_c$ it is found that $\tau\sim(1-2)\times10^{-14}$~s.\cite{Cherpak13,Barannik13} Thus, OP-BaFeCoAs may be closer to the dirty limit, for which $H_{c2}(\theta)$ may be expressed as\cite{Gurevich03}
\begin{equation} 
\label{GurevichHc2}
H_{c2}(\theta)\propto\frac{T-T_c}{a_1D_1(\theta) + a_2D_2(\theta)},
\end{equation}
with $a_{1,2} = 1 \pm \lambda_-/\lambda_0$, where $\lambda_- = \lambda_{11}-\lambda_{22}$, $\lambda_0 = (\lambda_-^2 + 4\lambda_{12}\lambda_{21})^{1/2}$, and $\lambda_{\alpha\beta}$ are the superconducing intra- ($\alpha=\beta$) and inter- ($\alpha\neq\beta$) band couplings. The angular dependency is contained in 
\begin{equation}
D_m(\theta) = \sqrt{D_m^{a^2}\cos^2\theta + D_m^aD_m^c\sin^2\theta},
\label{refeq8}
\end{equation}
being $D_m^{a, c}$ the electron diffusivities of band $m$ in the $a$ and $c$ directions. Normalizing equation~(\ref{GurevichHc2}) by $H_{c_2}(\theta=0)$, we obtain
\begin{equation}
\frac{H_{c2}(\theta)}{H_{c2}(0)} = \frac{a_1D_1^a + a_2D_2^a}{a_1D_1(\theta) + a_2D_2(\theta)}.
\label{refeq9}
\end{equation}
Defining $\delta \equiv a_2D_2^a/a_1D_1^a$ (that represents the relative contribution of the second band), and $\gamma_m=\sqrt{D_m^a/D_m^c}$, the anisotropy of each band, it follows that
\begin{equation}\label{GurevichFinal}
\frac{H_{c_2}(\theta)}{H_{c_2}(0)}  = \frac{1 + \delta}{\sqrt{\cos^2\theta+\gamma^{-2}_1\sin^2\theta} + \delta\sqrt{\cos^2\theta+\gamma^{-2}_2\sin^2\theta}}.
\end{equation}
The line in Fig.~\ref{reffig3}\textbf{d} is the best fit of equation~(\ref{GurevichFinal}) to the $H_{c2}(\theta)$ data resulting from the analysis of $\Delta\sigma$. As it can be seen, the agreement is excellent in the entire $\theta$ range, and leads to $\delta= 0.61\pm0.21$, $\gamma_1= 8.7\pm2.2$, $\gamma_2= 1.28\pm0.16$. As OP-BaFeCoAs is not strictly in the dirty limit these values may be just approximated, but the result suggests that both bands contribute similarly, and that the observed anisotropy comes essentially from one of the bands. 

\subsection*{Comparison with the usual procedure to obtain the upper critical field}

\begin{figure}[ht]
\centering
\includegraphics[width=0.68\linewidth]{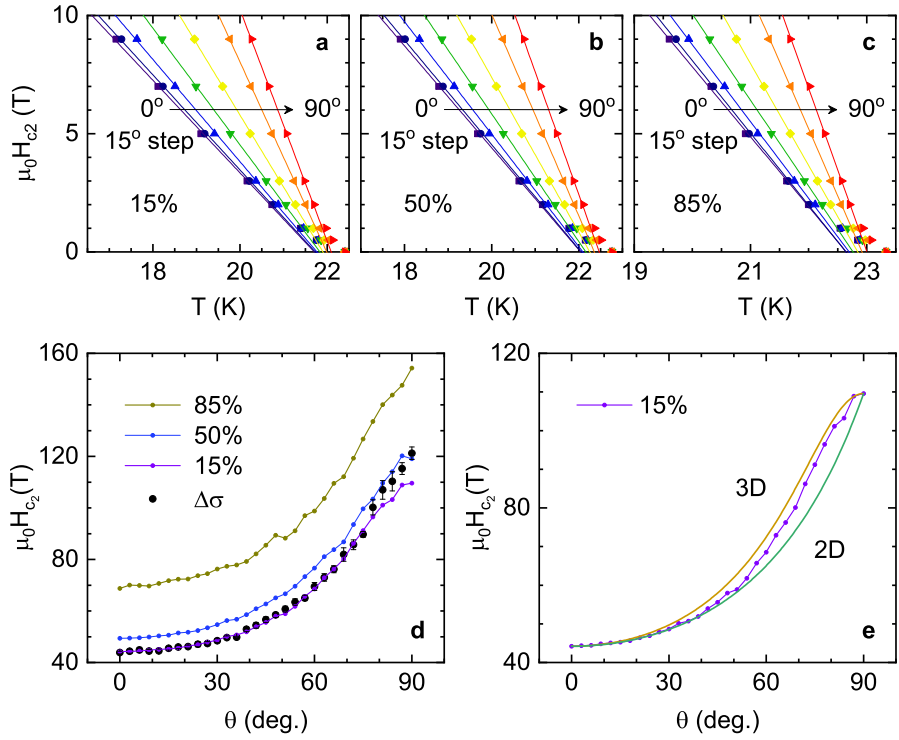}
\caption{\textbf{a-c} $H_{c2}(T)_\theta$ calculated as the magnetic field and temperature pairs at which the resistivity falls to a given percentage of the background resistivity at $T_c$. The behavior is linear up to very close to $T_c$ (within the resistive transition width), where a $T_c$ distribution may strongly affect the $\rho(T)_{H,\theta}$ behavior. The extrapolation to 0~K is presented in \textbf{d}, where it is shown that no criteria match the result from the $\Delta\sigma$ analysis (black circles), which shows a more pronounced maximum close to $90^\circ$. In \textbf{e}, it is shown that the $H_{c2}$ resulting from a criteria neither follows the 3D-aGL angular dependence.}
\label{reffig4}
\end{figure}

Figures~\ref{reffig4}\textbf{a-c} shows $H_{c2}(T)_\theta$, as obtained from the magnetic field and temperature pairs at which the resistivity $\rho(T)_{H,\theta}$ falls to a given percentage of the background resistivity, which is the most often used procedure in the literature to determine the upper critical field. As it can be seen, the obtained $H_{c2}(T)_\theta$ is linear with $T$, except very close to $T_c$ (within the resistive transition width). The linear extrapolation of $H_{c2}$ to $T=0$~K is presented in Fig.~\ref{reffig4}\textbf{d} as a function of $\theta$. As expected in view of the important fluctuation effects around $T_c$, the $H_{c2}$ amplitude is highly dependent on the chosen criterion. Furthermore, the $H_{c2}(\theta)$ profile is different from the one resulting from the $\Delta\sigma$ analysis (solid data points), with a less pronounced maximum near $\theta=90^\circ$. However, as shown in Fig.~\ref{reffig4}\textbf{e}, the calculated $H_{c2}(\theta)$ resulting from a criterion neither follows the behavior predicted by the 3D anisotropic GL approach, which is consistent with previous works.\cite{Murphy13,Hanisch15}

\subsection*{3D-anisotropic GL scaling of the resistivity around $T_c$}

\begin{figure}[ht]
\centering
\includegraphics[width=\linewidth]{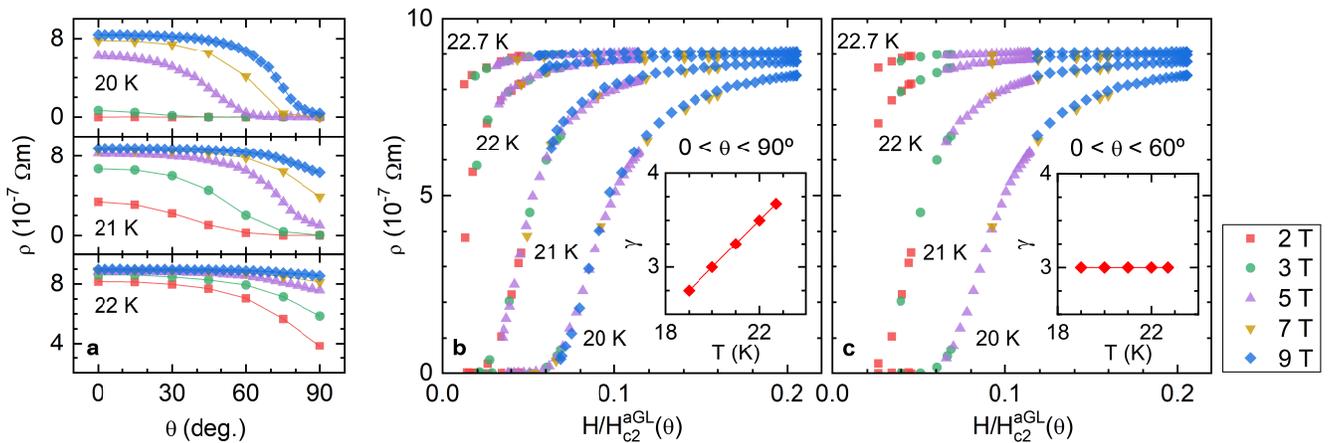}
\caption{Analysis of the 3D-aGL scaling of the resistivity for different temperatures near $T_c$. \textbf{a} Some examples of the raw unscaled data. \textbf{b} Scaling obtained by using a temperature dependent anisotropy factor (shown in the inset). \textbf{c} Scaling of the data with $\theta<60^\circ$ (for which $H_{c2}$ still follows the 3D-aGL approach). In the latter, an excellent scaling is obtained with a temperature independent $\gamma$.}
\label{reffig5}
\end{figure}

Previous works in different FeSC families showed that the $\rho(T)_{H,\theta}$ data scale when represented against $H(\cos^2\theta+\sin^2\theta/\gamma^2)^{1/2}$, according to the 3D-aGL approach.\cite{Wang08,Yi16,Yuan15,Jia09,Kalenyuk17} This scaling was used to determine the anisotropy factor, that in most cases was found to be temperature-dependent.\cite{Wang08,Yi16,Yuan15,Jia09} As it can be seen in Fig.~\ref{reffig5}\textbf{b} such scaling also works with our data (examples of unscaled data for 20-22~K are presented in Fig.~\ref{reffig5}\textbf{a}). The resulting anisotropy factor (inset in Fig.~\ref{reffig5}\textbf{b}) presents a significant temperature dependence, increasing linearly from 2.75 at 19~K to 3.75 near $T_c$. Nonetheless, this $\gamma$-dependence must be affected to some extent by the anomalous $H_{c2}(\theta)$ observed above. To test this hypothesis, we repeated the scaling only using data with $\theta<60^\circ$, for which the $H_{c2}(\theta)$ is still close to the \textit{single-band} 3D-aGL prediction, as can be seen in Fig.~\ref{reffig5}\textbf{c}. With this criterion, a good scaling is achieved with a temperature-independent anisotropy factor $\gamma=3$ (a similar agreement is obtained with a $\gamma$ value between 2.7 and 3.3). A $T$-independent $\gamma$ was also found in Ref.~\citeonline{Kalenyuk17} in Ba$_{1-x}$Na$_x$Fe$_2$As$_2$ ($x=0.35-0.4$), after excluding $\theta$ data close to 90$^\circ$. The failure of the scaling at high $\theta$ in the aforementioned paper was attributed to a transition to a 2D behavior. Nevertheless, in the present case such a possibility is ruled out by the excellent agreement of the 3D approach for $\Delta\sigma$ with the experimental data up to $\theta=90^\circ$.

\subsection*{Analysis of the irreversibility field}

\begin{figure}[ht]
\centering
\includegraphics[width=0.6\linewidth]{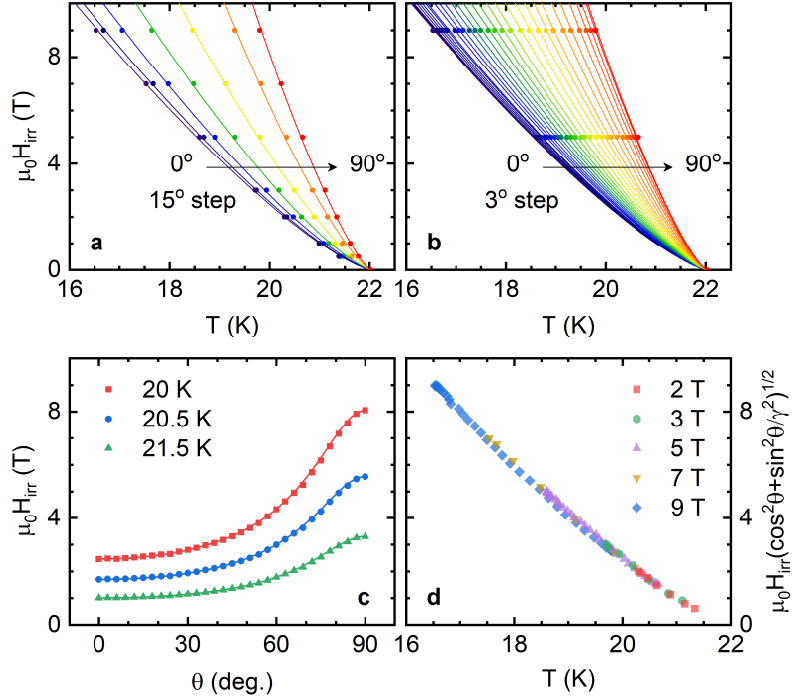}
\caption{\textbf{a} Temperature dependence of the irreversibility field for different field orientations, obtained  from a 1\% of $\rho_B(T_c)$ criterion. The lines are a fit of a power law $H_{irr}=A(\theta)(T-T_c)^n$ to the entire set of data, that leads to $n\approx1.3$. \textbf{b} $H_{irr}(T)_\theta$ in steps of 3$^\circ$, obtained from the detailed 5~T and 9~T data in Figs.~\ref{reffig1}\textbf{f, h}. The lines are fits to the above power law with $n$ fixed to 1.3, that allowed us to obtain the $H_{irr}(\theta)$ presented in \textbf{c}. In this case the 3D-aGL approach (solid line) accounts for the angular dependence of $H_{irr}$ in the entire $\theta$ range. \textbf{d} 3D-aGL scaling of the irreversibility field, obtained with $\gamma=3.27$.}
\label{reffig6}
\end{figure}

For completeness, we present the temperature dependence of the irreversibility field, $H_{irr}$, for different $\theta$ values as shown in Fig.~\ref{reffig6}\textbf{a}. $H_{irr}$ was determined by using a 1\% criterion on $\rho_B(T_c)$. The solid lines were obtained as the best fit to the power law $H_{irr} = A(\theta)(T_c-T)^n$ to all $H_{irr}(T)_{\theta}$ curves by leaving $A$ as a free parameter for each $\theta$ and the same $n$ for all curves. The fit quality is excellent, and leads to $n = 1.30 \pm 0.14$, close to the value found in other iron-based superconductors,\cite{Bendele10,Prando11b} and in high-$T_c$ cuprates.\cite{Malozemoff88,Yeshurun88}

The 5 T and 9 T series, for which $H_{irr}$ was obtained in $\theta$-steps of $3^\circ$, were analyzed to obtain the angular dependence of the irreversibility field, as shown in Fig.~\ref{reffig6}\textbf{b}. The solid lines in this figure are fits to the previously mentioned power law with $n = 1.30$ and $A$ as the only free parameter for each $\theta$-series. From these curves $H_{irr}(\theta)$ was obtained for different temperatures (see Fig.~\ref{reffig6}\textbf{c}). Contrary to the results obtained for the upper critical field, we found that the irreversibility field follows the 3D-aGL angular dependence closely (equation~(\ref{refeq1}), solid lines). This result is confirmed by the excellent 3D-aGL scaling presented in Fig.~\ref{reffig6}\textbf{d}, that was obtained with $\gamma=H_{irr}^\parallel/H_{irr}^\perp=3.27$ (consistent with the value obtained in the previous section). We speculate that this discrepancy may arise from the vortex pinning by defects not being appreciably affected by the multiband electronic structure.

\section*{Conclusions}


The electrical resistivity was measured in a high-quality OP-BaFeCoAs crystal, under magnetic fields with different amplitudes and orientations relative to the crystal $c$ axis. The rounding observed just above $T_c(H)$ was interpreted in terms of Cooper pairs created by thermal fluctuations. The comparison with a generalization to finite fields of the AL approach for fluctuation effects, allowed a criterion-independent determination of $H_{c2}(\theta)$ to be made. The result differs significantly with the prediction of the single-band 3D-anisotropic Ginzburg-Landau approach, particularly for magnetic fields close to the crystal $ab$ layers. The behavior is similar to the one of quasi-2D superconductors, but this possibility is inconsistent with the 3D nature of the superconducting fluctuations above $T_c(H)$. $H_{c2}(\theta)$ was then successfully compared with a theoretical approach for dirty two-band superconductors. Although OP-BaFeCoAs is not strictly in the dirty limit, the result suggests that both bands contribute with roughly the same weight, and that the observed anisotropy comes essentially from a highly anisotropic band ($\gamma_1= 8.7\pm2.2$), while the other band is almost isotropic ($\gamma_2= 1.28\pm0.16$). This result contrasts with alternative explanations for a similar anomalous $H_{c2}(\theta)$ behavior observed in these materials.\cite{Su14,Hao15,Murphy13,Hanisch15}

We have also found that the resistivity scales with $H(\cos^2\theta+\sin^2\theta/\gamma^2)^{1/2}$, as predicted by the 3D-aGL approach, if the data are restricted to $\theta<60^\circ$, where $H_{c2}(\theta)$ is reasonably well described by the 3D-aGL expression (equation~(\ref{refeq1})). This leads to a temperature-independent {\it effective} $\gamma$, in striking contrast with previous works reporting a strongly temperature-dependent $\gamma$ near $T_c$.\cite{Wang08,Yi16,Yuan15,Jia09}
Finally, in contrast with the upper critical field, the irreversibility field (determined from a 1\% criterion on the normal-state resistivity) presents an angular dependence fully consistent with the one expected for 3D anisotropic superconductors, suggesting that the multiband electronic structure does not noticeably affect the vortex pinning. Nevertheless, it is possible that the symmetry of the vortex lattice in these materials could be affected by the presence of several bands, as recently observed in MgB$_2$.\cite{Curran15}

It would be interesting to extend the present study to other FeSC families (e.g., 112,\cite{Katayama13} 10-3-8 and 10-4-8,\cite{Kakiya11,Lohnert11,Ni11}  
for which a possible quasi-2D behavior\cite{Sonora17} may also affect the $H_{c2}(\theta)$ angular dependence), and to probe signatures of crossband pairing, that could be present in these materials, as it has been recently proposed.\cite{Vargas20}

\section*{Methods}

The Ba(Fe$_{1-x}$Co$_{x}$)$_2$As$_2$ ($x=0.065$) crystal was grown following the procedure described in previous works.\cite{Sefat08,Sefat09} It is a 2.902~mg plate with a 3.1~mm$^2$ surface parallel to the crystal $ab$ layers, and a thickness of 144~$\upmu$m along the crystal $c$ axis (as determined from the density calculated from the lattice parameters).

\begin{figure}[ht]
\centering
\includegraphics[width=0.6\linewidth]{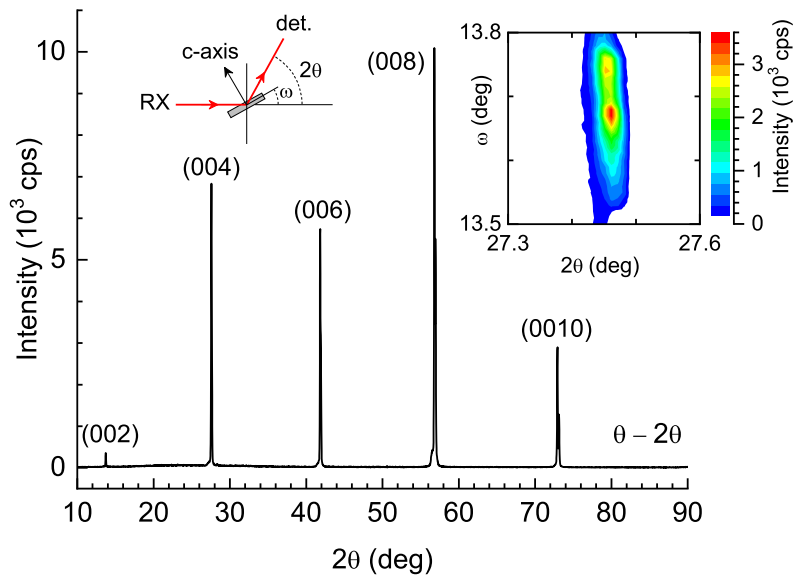}
\caption{X-ray diffraction pattern obtained with the geometry indicated in the diagram. Only the (00$l$) reflections are observed. Inset: $\omega-2\theta$ intensity map for the (004) peak, showing that the dispersion in the orientation of the crystal $c$ axis is about $0.2^\circ$.}
\label{reffig7}
\end{figure}

The homogeneity of the crystal structure was tested by $x$-ray diffraction. As it can be seen in Fig.~\ref{reffig7}, the $\theta-2\theta$ pattern (performed with a Rigaku Miniflex II diffractometer with a Cu target and a graphite monochromator) presents only $(00l)$ reflections, indicating the excellent structural quality of the crystal. The resulting $c$ axis lattice parameter (that is the same as the FeAs layers interdistance, $s$) is 12.979(2) \r{A}, in agreement with data in the literature for crystals with a similar composition.\cite{Sefat09,Ni08} The inset in Fig.~\ref{reffig7} represents the $\omega-2\theta$ intensity map for the (004) peak, performed with a Panalytical-Empyrean diffractometer. As it can be seen, the dispersion in $\omega$ is $\sim0.2^\circ$, which indicates the excellent alignment of the crystal $c$ axis. 

The $ab$ layers dc resistivity $\rho$ was measured in the presence of magnetic fields up to 9 T with different orientations $\theta$ relative to the crystal $c$ axis. To obtain these measurements, a Quantum Design's Physical Property Measurement System (PPMS) equipped with a rotating sample holder with an angular resolution of about $0.01^\circ$, was used. 
The electrical contacts (in-line configuration) were made with four gold wires (50 $\mu$m diameter) attached to the crystal with silver paste. The excitation current was 1 mA. To avoid the mechanical backlash, the target angles were always approached from an angle smaller by a few degrees. Prior to the measurements, the precise $\theta=90^\circ$ position was identified by a $\rho(\theta)$ calibration measurement at 20~K under a 9~T magnetic field (see the inset in Fig.~\ref{reffig1}). The actual $\theta=90^\circ$ position was found to be $\sim2^\circ$ away from the nominal value, probably due to the General Electric varnish used to attach the sample to the holder.


\section*{Acknowledgements}

This work was supported by the Agencia Estatal de Investigaci\'on (AEI) and Fondo Europeo de Desarrollo Regional (FEDER) through projects FIS2016-79109-P and PID2019-104296GB-I00, and by Xunta de Galicia (grant GRC no. ED431C 2018/11). The work at Oak Ridge National Laboratory was funded by U.S. Department of Energy, Materials Sciences and Engineering Division, Basic energy Sciences. SSS acknowledges support from CNPq. I.F. Llovo acknowledges financial support from Xunta de Galicia through grant ED481A-2020/149. Authors would like to thank the use of RIAIDT-USC analytical facilities.

\section*{Author contributions statement}

J.M. conceived the experiments, A.S.S. fabricated the sample, J.M. and J.P. conducted the experiments, I.F.L., C.C., and J.M. analyzed the results, D.S., A.P., and S.S. also helped in data analysis, J.M. and  I.F.L. wrote the manuscript. All authors reviewed the manuscript. 

\section*{Additional information}

\textbf{Competing interests} The authors declare no competing interests. 

\newpage

\end{document}